\documentclass[cameraready]{Interspeech} % Camera-ready

\usepackage{booktabs}
\usepackage{graphicx}
\usepackage{multirow}
\usepackage{array}

\title{VoxENES 2026: Benchmarking Generalization of Speech Spoofing Detectors Against LLM-Era TTS and Voice Conversion}

\author[]{Aastha}{Sharma}
\author[]{Guangjing}{Wang}
\address{University of South Florida, Tampa, FL, USA
}
\email{aasthasharma@usf.edu, guangjingwang@usf.edu}


\begin{document}

\maketitle

\begin{abstract}
Modern LLM-driven text-to-speech (TTS) and voice conversion (VC) systems produce synthetic speech that differs from the generators represented in many legacy spoofing benchmarks. This mismatch creates a temporal generalization gap that can overestimate detector robustness under real-world post-processing conditions. We bridge this gap by introducing VoxENES 2026, a bilingual (English and Spanish) benchmark of 53,628 audio samples generated using 10 contemporary speech synthesis methods and evaluated under 10 standardized post-processing conditions. Using VoxENES 2026, we benchmark eight pretrained detectors without fine-tuning and observe substantial performance degradation: the best model achieves 28.98\% EER overall, while most perform near or below random chance across modern generators and perturbations. Our results highlight the reliance on brittle artifacts in current detectors and establish VoxENES 2026 as a practical testbed for developing robust audio spoofing countermeasures.
\end{abstract}

\section{Introduction}

Robust speech spoofing and deepfake detection are essential to preserve trust in speech-based authentication~\cite{wang2025clearmask, guo2024wavepurifier, guo2023phantomsound, wang2023vsmask}. This need is growing as voice becomes a biometric and a control channel for speech-driven agents and assistive technologies. If synthetic speech becomes indistinguishable from genuine speech, detection failures not only compromise security protocols but also erode trust in audio evidence.

Many benchmark datasets are proposed for speech spoofing and deepfake detection evaluation. For example, the ASVspoof challenge series has been the primary driver of spoofing countermeasure development. ASVspoof 2019~\cite{todisco2019asvspoof} introduces logical access (LA) with TTS and VC, and physical access (PA) with replay tracks. ASVspoof 2021~\cite{yamagishi2021asvspoof} adds the deepfake task targeting compressed manipulated speech, and ASVspoof 5~\cite{wang2026asvspoof5} introduces crowdsourced data with adversarial attacks at scale. In addition to ASVspoof, WaveFake~\cite{frank2021wavefake} provides a multilingual dataset from six neural vocoder architectures. The In-the-Wild dataset~\cite{muller2022inthewild} includes real-world deepfakes of celebrities. The MLAAD~\cite{muller2024mlaad} dataset expands coverage to 23 languages and 54 TTS models. The VoiceWukong~\cite{yan2024voicewukong} benchmarks 12 detectors against 34 commercial and open-source tools with post-processing manipulations.

Yet, existing benchmarks primarily rely on speech synthesis systems before 2024 and fail to capture artifact patterns produced by the modern large language model (LLM)-driven generation pipelines. For example, text-to-speech (TTS) designs include autoregressive language-model-based synthesis, such as VoxCPM~\cite{zhou2025voxcpm} and Qwen3-TTS~\cite{hu2026qwen3}; flow-matching models, including GLM-TTS~\cite{cui2025glm}, CosyVoice 3~\cite{du2025cosyvoice3}, and Chatterbox~\cite{resemble2025chatterbox}; diffusion-based systems like FlashLabs Chroma~\cite{chen2026flashlabs}; and hybrid DiT architectures, such as VibeVoice~\cite{pengvibevoice}. For voice conversion (VC), zero-shot approaches such as Seed-VC~\cite{liu2024seedvc}, tone-color extraction methods such as OpenVoice v2~\cite{myshell2024openvoicev2}, and retrieval-based systems like RVC v2~\cite{rvcproject2024rvcwebui} have substantially improved naturalness and speaker similarity.

Evaluation on temporally stale benchmarks can overestimate real-world robustness as speech synthesis techniques improve. LLM-driven TTS and VC systems produce synthetic audio with acoustic characteristics that differ substantially from earlier spoofing corpora, creating a \emph{data drifting issue} for existing detectors. Deepfake detection models that perform well on older benchmarks may fail when deployed against newer deepfake generators. This mismatch reflects a cat-and-mouse dynamic in which detectors learn cues tied to previous synthesis artifacts, while generators and post-processing steps progressively suppress or conceal those cues. As a consequence, the rapid LLM-driven TTS and VC evolution motivates updated evaluation benchmarks.

To study the generalization ability of deepfake detectors under the data drifting issue in the era of LLM, we introduce a modern bilingual benchmark dataset \textbf{VoxENES 2026}. We generate synthetic audios from 10 LLM-driven synthesis methods (7 TTS, 3 VC), covering English and Spanish audios. In addition, real-world audios are frequently subjected to transformations such as compression, noise, and resampling, which can substantially affect detector behavior~\cite{yamagishi2021asvspoof,yan2024voicewukong,muller2024mlaad}.
Therefore, VoxENES 2026 explicitly models the post-processing setting by applying a standardized set of audio augmentation methods that simulate common transmission and manipulation effects, including codec compression, additive noise, resampling, speed perturbation, and loudness normalization. 

With VoxENES 2026, we evaluate eight pretrained deepfake detectors without fine-tuning to measure out-of-distribution generalization. Our results show that existing detectors suffer from substantial degradation relative to reported performance on legacy benchmarks. The best detector achieves only 28.98\% EER overall, while most detectors perform near or below random chance. These findings suggest that many current detectors rely on brittle artifact cues that may not generalize across new TTS and VC generations and realistic post-processing. By providing a modern benchmark and a controlled out-of-distribution evaluation, our work establishes a necessary testbed for measuring real-world speech spoofing detector robustness and for driving detector development that keeps pace with the evolving synthesis frontier.

In summary, the main contributions of our work are:
\begin{itemize}
    \item We introduce a modern bilingual benchmark dataset \textbf{VoxENES 2026} using LLM-driven TTS and VC systems with realistic post-processing to emulate deployment-time distribution shifts, totaling 53,628 audio samples available at \url{https://www.kaggle.com/datasets/interspeech2712/voxenes-2026}.
    \item We benchmark eight pretrained detectors and reveal a substantial generalization gap, exposing detector-specific blind spots and robustness failures across synthesis methods.
\end{itemize}

\section{VoxENES 2026}

VoxENES 2026 consists of three components: (1) real speech from established corpora, (2) synthetic audio generated by modern TTS and VC systems, and (3) post-processed augmented variants simulating real-world transmission conditions. As shown in Table~\ref{tab:dataset_summary_a}, VoxENES includes 3,028 real speech samples and 50,600 synthetic samples (original synthetic 4,600 and augmented synthetic 46,000).

\subsection{Real Speech and Standardization}
As shown in Table~\ref{tab:dataset_summary_b}, we drew 1,500 real English speech samples from 40 speakers from LibriSpeech~\cite{panayotov2015librispeech}. We drew 1,528 real Spanish speech samples from 96 speakers from VoxPopuli~\cite{wang2021voxpopuli}. All audio was standardized to 16~kHz mono WAV. For detector compatibility, samples were capped at 4 seconds with truncation and zero-padding. This fixed-length standardization was applied uniformly to both bonafide and synthetic samples so that any padding-induced cues are shared across classes rather than correlated with the spoofing label. We, therefore, do not expect zero-padding to serve as a discriminative shortcut, though a detailed analysis of padding artifacts is left to future work. Note that for the downstream voice conversion tasks, the people whose voices appear in the real speech samples are not used as target speakers.

\begin{table}[t]
\caption{VoxENES 2026 dataset summary.}
\label{tab:dataset_summary_a}
\centering
\small
\begin{tabular}{l r l}
\toprule
\textbf{Component} & \textbf{Count} & \textbf{Notes} \\
\midrule
Real speech & 3,028 & LibriSpeech + VoxPopuli \\
Original synthetic & 4,600 & TTS + VC originals \\
Augmented synthetic & 46,000 & 10$\times$ post-processing \\
\midrule
\textbf{Total} & \textbf{53,628} & EN + ES combined \\
\bottomrule
\end{tabular}
\end{table}

\begin{table}[t]
\caption{Language and source breakdown in VoxENES 2026.}
\label{tab:dataset_summary_b}
\centering
\small
\begin{tabular}{l r r r}
\toprule
\textbf{Category} & \textbf{EN} & \textbf{ES} & \textbf{Total} \\
\midrule
Real speech & 1,500 & 1,528 & 3,028 \\
TTS (7 methods) & 11,000 & 6,600 & 17,600 \\
VC (3 methods) & 16,500 & 16,500 & 33,000 \\
\midrule
\textbf{Total} & \textbf{29,000} & \textbf{24,628} & \textbf{53,628} \\
\bottomrule
\end{tabular}
\end{table}

\subsection{Synthesis Systems}
We selected seven TTS and three VC systems representing state-of-the-art LLM-driven synthesis techniques across diverse architectures as shown in Table~\ref{tab:synthesis_systems}. The TTS systems include: (i) VoxCPM 1.5~\cite{zhou2025voxcpm}, an autoregressive language model from HKUST; (ii) Qwen3-TTS~\cite{hu2026qwen3}, a streaming LM from Alibaba with native multilingual support; (iii) GLM-TTS~\cite{cui2025glm}, a flow-matching model from Zhipu AI; (iv) FlashLabs Chroma~\cite{chen2026flashlabs}, a diffusion-based system; (v) VibeVoice~\cite{pengvibevoice}, combining DiT with flow matching; (vi) CosyVoice 3~\cite{du2025cosyvoice3}, an LM plus flow-matching system from Alibaba; and (vii) Chatterbox ML~\cite{resemble2025chatterbox}, a flow-matching model from Resemble AI. The three VC systems represent distinct paradigms: (i) Seed-VC~\cite{liu2024seedvc} uses diffusion-based zero-shot conversion with Whisper and WavLM features; (ii) OpenVoice v2~\cite{myshell2024openvoicev2} performs tone-color extraction and transfer; and (iii) RVC v2~\cite{rvcproject2024rvcwebui} uses HuBERT-based retrieval with HiFi-GAN vocoding. All TTS systems were released or updated in 2025, ensuring they represent modern threats that older detection models have never encountered. VC samples are generated using disjoint source audio partitions and multiple target speaker references to maintain diversity.

\begin{table*}[!t]
\caption{TTS and VC systems used in VoxENES 2026.}
\label{tab:synthesis_systems}
\centering
\begin{tabular}{l l l l r r r r l}
\toprule
\textbf{System} & \textbf{Type} & \textbf{Architecture} & \textbf{Developer} & \textbf{EN} & \textbf{ES} & \textbf{Total} & \textbf{Year} & \textbf{ES Mode} \\
\midrule
VoxCPM 1.5~\cite{zhou2025voxcpm} & TTS & Autoregressive LM & HKUST & 200 & -- & 200 & 2025 & -- \\
Qwen3-TTS~\cite{hu2026qwen3} & TTS & Streaming LM & Alibaba & 200 & 200 & 400 & 2025 & Native \\
GLM-TTS~\cite{cui2025glm} & TTS & Flow matching & Zhipu AI & 200 & -- & 200 & 2025 & -- \\
FlashLabs Chroma~\cite{chen2026flashlabs} & TTS & Diffusion & FlashLabs & 200 & -- & 200 & 2025 & -- \\
VibeVoice~\cite{pengvibevoice} & TTS & DiT + flow matching & Vibe AI & 200 & -- & 200 & 2025 & -- \\
CosyVoice 3~\cite{du2025cosyvoice3} & TTS & LM + flow matching & Alibaba & -- & 200 & 200 & 2025 & Native \\
Chatterbox ML~\cite{resemble2025chatterbox} & TTS & Flow matching & Resemble AI & -- & 200 & 200 & 2025 & Native \\
\midrule
Seed-VC~\cite{liu2024seedvc} & VC & Diffusion + zero-shot & ByteDance & 500 & 500 & 1,000 & 2025 & -- \\
OpenVoice v2~\cite{myshell2024openvoicev2} & VC & Tone cloning & MyShell AI & 500 & 500 & 1,000 & 2024 & -- \\
RVC v2~\cite{rvcproject2024rvcwebui} & VC & Retrieval + HiFi-GAN & RVC-Project & 500 & 500 & 1,000 & 2023 & -- \\
\bottomrule
\end{tabular}
\end{table*}

\begin{table}[h!]
\caption{Post-processing augmentation techniques.}
\label{tab:augmentations}
\centering
\scalebox{0.86}{
\begin{tabular}{l l}
\toprule
\textbf{Augmentation} & \textbf{Description} \\
\midrule
mp3\_64k & MP3 encoding at 64 kbps \\
aac\_128k & AAC encoding at 128 kbps \\
noise\_white\_10db & White Gaussian noise at 10 dB SNR \\
noise\_white\_20db & White Gaussian noise at 20 dB SNR \\
noise\_babble\_15db & Multi-speaker babble noise at 15 dB SNR \\
resample\_8k & Downsample to 8 kHz, upsample to 16 kHz \\
resample\_16k & Resample to 16 kHz (control) \\
speed\_fast & 1.1$\times$ playback speed \\
speed\_slow & 0.9$\times$ playback speed \\
volume\_norm & Peak normalization to $-$3 dBFS \\
\bottomrule
\end{tabular}
}
\end{table}

\subsection{Post-Processing Augmentations}
Each original synthetic sample was subjected to one post-processing operation in Table~\ref{tab:augmentations} that reflect common transmission and editing effects: lossy codec compression (MP3 at 64 kbps and AAC at 128 kbps), additive noise perturbations (white noise at 10/20 dB SNR and babble noise at 15 dB SNR), bandwidth and sampling-rate distortions (downsampling to 8 kHz with restoration to 16 kHz, plus a 16 kHz resampling control), playback-rate changes (1.1$\times$ and 0.9$\times$ speed), and amplitude normalization (peak normalization to $-$3 dBFS). This pipeline expands the synthetic subset and enables systematic robustness analysis across realistic post-processing conditions.

\section{Evaluation Setup}

\subsection{Detection Baselines}
We evaluated eight pretrained audio deepfake detection systems spanning graph neural networks, raw-waveform CNNs, self-supervised learning (SSL) models, transformer-based detectors, and speaker-embedding anomaly detection as shown in Table~\ref{tab:detectors}. None were retrained or fine-tuned on VoxENES 2026, enabling an honest temporal generalization evaluation. The detectors differ in their training corpora (e.g., ASVspoof 2019 LA, ASVspoof 2021 DF, ASVspoof 5, and VoxCeleb), so absolute EER values should be read as out-of-distribution generalization indicators rather than as a strictly controlled head-to-head comparison; the most directly comparable cases are detectors sharing the same training source.

\begin{table}[h]
\caption{Detection baselines evaluated on VoxENES 2026.}
\label{tab:detectors}
\centering
\scriptsize
\begin{tabular}{l l l}
\toprule
\textbf{Detector} & \textbf{Family} & \textbf{Training Data} \\
\midrule
AASIST2~\cite{jung2022aasist} & Graph NN & ASVspoof 2019 LA \\
RawNet2~\cite{tak2021endtoend} & Raw waveform CNN & ASVspoof 2021 DF \\
Wav2Vec2-AASIST & SSL + classifier & ASVspoof 2019 \\
Wav2Vec2-DF & SSL + classifier & Mixed deepfake \\
Wav2Vec2-Large & SSL + classifier & Mixed deepfake \\
AST-ASVspoof5~\cite{gong2021ast} & Spectrogram Transformer & ASVspoof 5 \\
Wav2Vec2-ASVspoof5 & SSL + classifier & ASVspoof 5 \\
ECAPA-TDNN~\cite{desplanques2020ecapa} & Speaker embeddings & VoxCeleb \\
\bottomrule
\end{tabular}
\end{table}

\begin{table*}[!h]
\caption{Per-augmentation EER (\%) across all detectors.}
\label{tab:per_aug_results}
\centering
\scalebox{0.88}{
\begin{tabular}{l c c c c c c c c}
\toprule
\textbf{Augmentation} & \textbf{AASIST2} & \textbf{RawNet2} & \textbf{W2V-AASIST} & \textbf{W2V-DF} & \textbf{W2V-Large} & \textbf{AST-ASV5} & \textbf{W2V-ASV5} & \textbf{ECAPA} \\
\midrule
original & 54.4 & 51.3 & 41.0 & 53.9 & 39.9 & 26.7 & 52.6 & 46.5 \\
aac\_128k & 53.1 & 50.2 & 42.5 & 53.6 & 37.9 & 31.8 & 51.5 & 46.9 \\
mp3\_64k & 53.7 & 50.1 & 42.5 & 54.1 & 39.8 & 48.4 & 52.0 & 47.2 \\
noise\_white\_10db & 67.5 & 27.9 & 32.3 & 69.5 & 61.2 & 17.4 & 52.5 & 33.3 \\
noise\_white\_20db & 59.0 & 40.0 & 44.5 & 68.9 & 63.3 & 18.3 & 49.8 & 36.2 \\
noise\_babble\_15db & 65.6 & 33.0 & 22.9 & 50.8 & 45.1 & 25.7 & 46.3 & 44.5 \\
resample\_8k & 59.1 & 55.6 & 37.5 & 42.0 & 34.3 & 29.5 & 52.1 & 46.4 \\
resample\_16k & 54.1 & 49.7 & 42.3 & 54.1 & 39.7 & 27.1 & 52.3 & 46.1 \\
speed\_fast & 56.0 & 53.4 & 38.6 & 52.2 & 37.5 & 26.1 & 59.0 & 38.5 \\
speed\_slow & 52.7 & 47.8 & 45.0 & 52.5 & 42.4 & 30.2 & 49.8 & 42.0 \\
volume\_norm & 56.7 & 47.1 & 42.0 & 53.9 & 40.1 & 28.5 & 52.2 & 46.6 \\
\bottomrule
\end{tabular}
}
\end{table*}

\begin{figure*}
    \centering
    \includegraphics[width=0.98\textwidth]{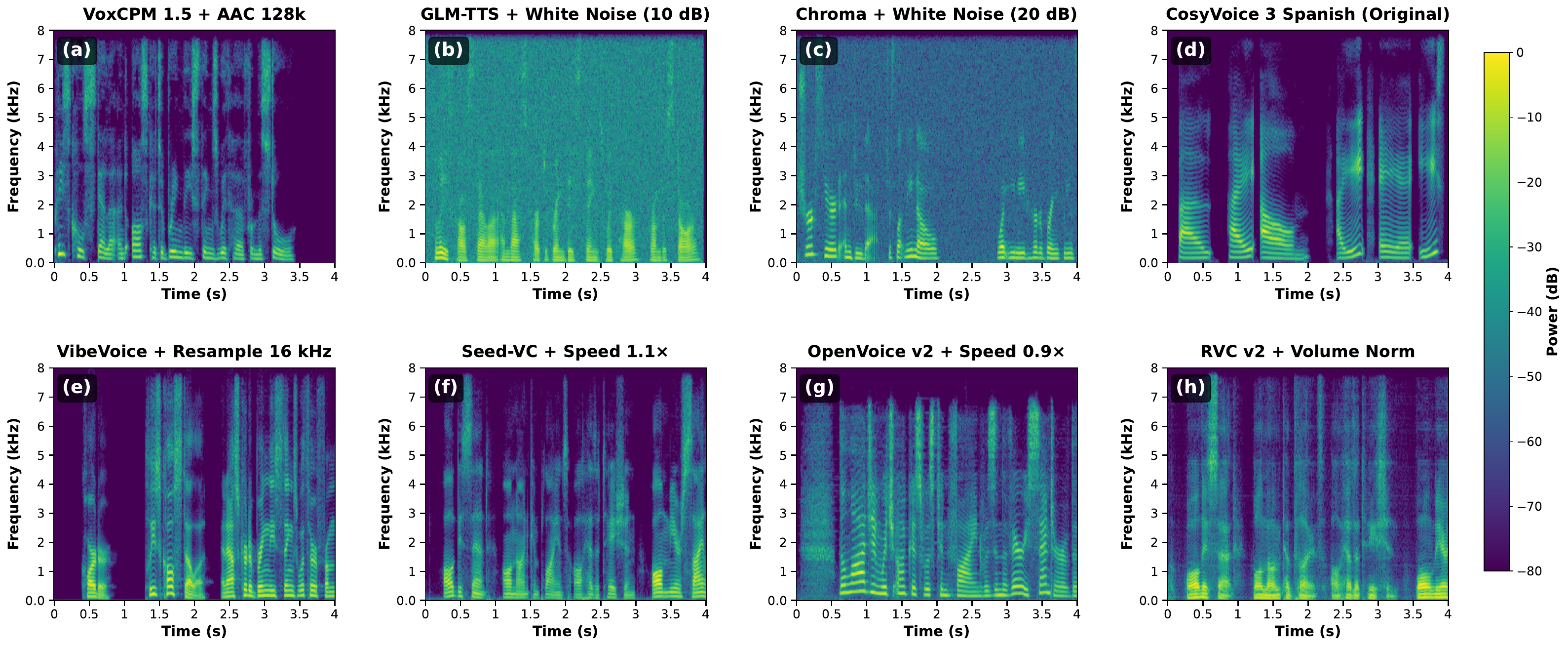}
    \caption{Representative spectrograms across multiple TTS and VC systems and augmentation conditions in VoxENES 2026.}
    \label{fig:multi_method_qualitative}
    % \vspace{-10pt}
\end{figure*}

\begin{figure}[h!]
    \centering
    \includegraphics[width=0.48\textwidth]{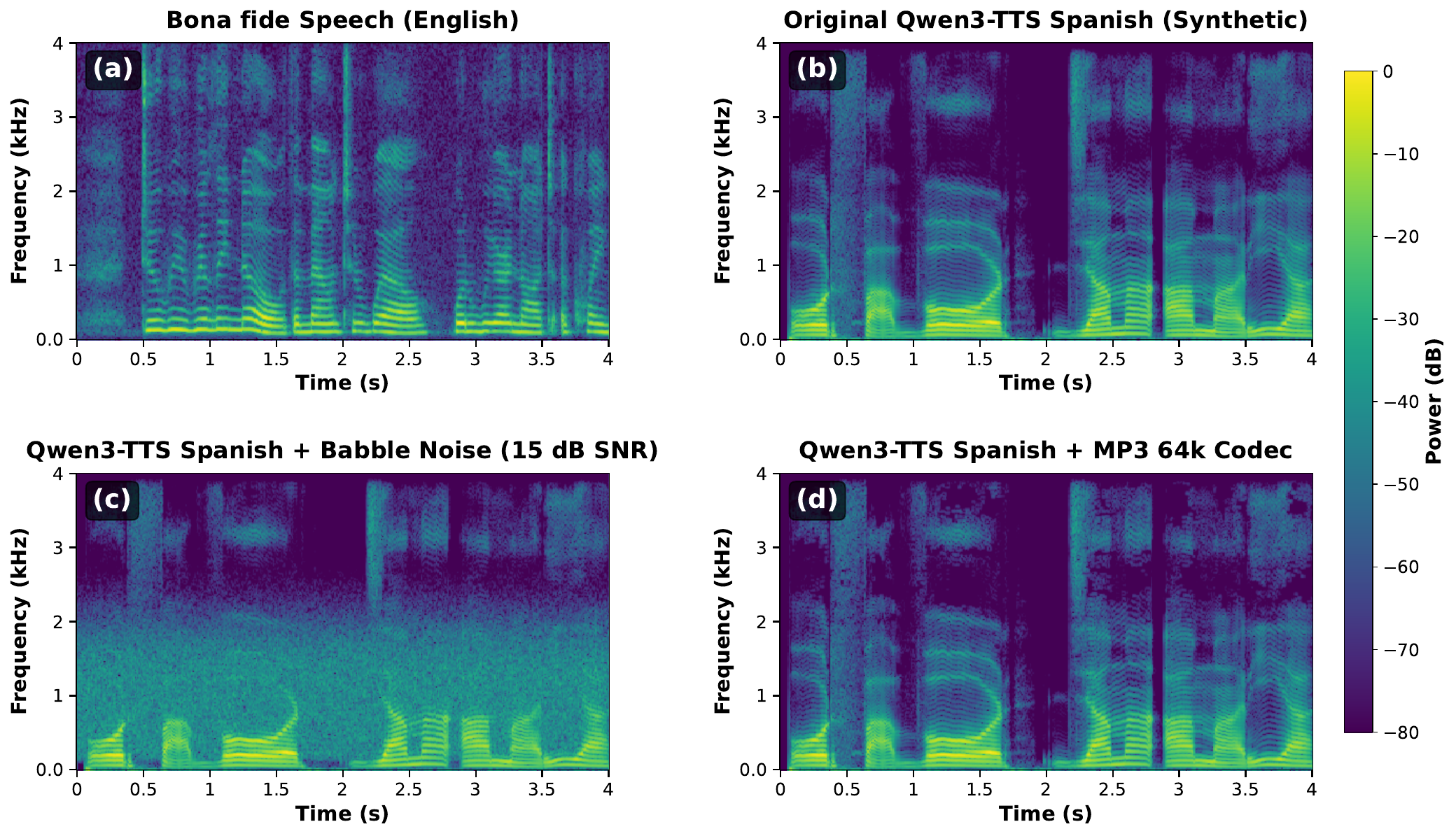}
    \caption{Qualitative spectrogram comparison.}
    \label{fig:qwen3_qualitative}
    \vspace{-10pt}
\end{figure}

All detectors were evaluated in inference-only mode using published pretrained weights. We report Equal Error Rate (EER) and accuracy in the results. EER was computed from the ROC operating point where the false acceptance rate (FAR) and the false rejection rate (FRR) are equal. In particular, ECAPA-TDNN~\cite{desplanques2020ecapa} is a speaker verification model rather than a dedicated spoof detector. Thus, we use an anomaly-based scoring scheme, where a reference centroid is computed from real speech audio, and the cosine distance from this centroid is used as the spoofing score. The system's performance is then evaluated using the EER, determined by the threshold at which the FAR and FRR are equivalent.

\section{Evaluation Results}

\subsection{Qualitative Spectrogram Analysis}

Figures~\ref{fig:multi_method_qualitative} and~\ref{fig:qwen3_qualitative} visualize the spectral variability across bonafide (real) speech, synthetic speech, and post-processed synthetic samples in VoxENES 2026. The examples highlight that realistic perturbations, such as additive noise and lossy codec compression, can reshape time-frequency structure and attenuate synthesis artifacts that many detectors implicitly rely on, providing a qualitative rationale for the observed performance changes under post-processing.

\subsection{Impact of Post-Processing}

As shown in Table~\ref{tab:per_aug_results}, post-processing impacts detector performance in a highly non-uniform and occasionally counterintuitive manner. Adding white noise reduces EER for AST-ASVspoof5 (from 26.7\% to 17.4\%) and RawNet2 (from 51.3\% to 27.9\%). This is plausibly because noise perturbs synthetic and bonafide speech differently and attenuates generator-specific artifacts, thereby inducing alternative discriminative cues. In contrast, MP3 compression degrades AST-ASVspoof5 (from 26.7\% to 48.4\% EER), consistent with a reliance on fine-grained spectral structure, particularly in higher frequencies, that is suppressed by lossy codec encoding.

\subsection{Overall Detection Performance}
\begin{table}[h!]
\caption{Overall detection results on VoxENES 2026.}
\label{tab:overall_results}
\centering
\scalebox{0.7}{
\begin{tabular}{l l c c c c}
\toprule
 \textbf{Detector} & \textbf{EER (\%)} & \textbf{Acc (\%)} & \textbf{TTS EER (\%)} & \textbf{VC EER (\%)} \\
\midrule
 AASIST2 & 57.86 & 42.13 & 61.2 & 49.9 \\
RawNet2 & 47.03  & 52.97 & 53.6 & 49.8 \\
Wav2Vec2-AASIST & 39.16  & 60.85 & 43.7 & 38.4 \\
Wav2Vec2-DF & 55.51 & 44.49 & 59.3 & 49.2 \\
Wav2Vec2-Large & 44.38 & 55.62 & 40.0 & 39.7 \\
AST-ASVspoof5 & \textbf{28.98} & \textbf{75.94} & \textbf{20.9} & \textbf{29.8} \\
Wav2Vec2-ASVspoof5 & 51.53 & 48.49 & 50.6 & 53.5 \\
ECAPA-TDNN & 43.22 & 56.78 & 28.5 & 52.5 \\
\bottomrule
\end{tabular}
}
\end{table}

As shown in Table~\ref{tab:overall_results}, among the evaluated models, only AST-ASVspoof5 achieves an EER below 30\% (i.e., 28.98\%), a result that remains insufficient for reliable field deployment. In addition, five of the eight detectors perform at or below the stochastic baseline (EER $\ge 47\%$), most notably AASIST2 (57.86\%), which exhibits inverted prediction behavior on this benchmark and suggests a severe domain mismatch between the training distribution and the VoxENES 2026 benchmark. This phenomenon occurs when a classifier learns dataset-specific artifacts, such as silence patterns or channel characteristics, that are reversed or absent in the target domain. Consequently, the model assigns higher bonafide scores to synthetic samples, misinterpreting spoofing cues as genuine speech features. Across models, TTS-generated samples are marginally easier to detect than VC-generated samples for some detectors, but the relative difficulty varies by detector and does not hold uniformly.

\subsection{Per-Method Detection Performance}

\begin{figure} [h!]
    \centering
\includegraphics[width=0.49\textwidth]{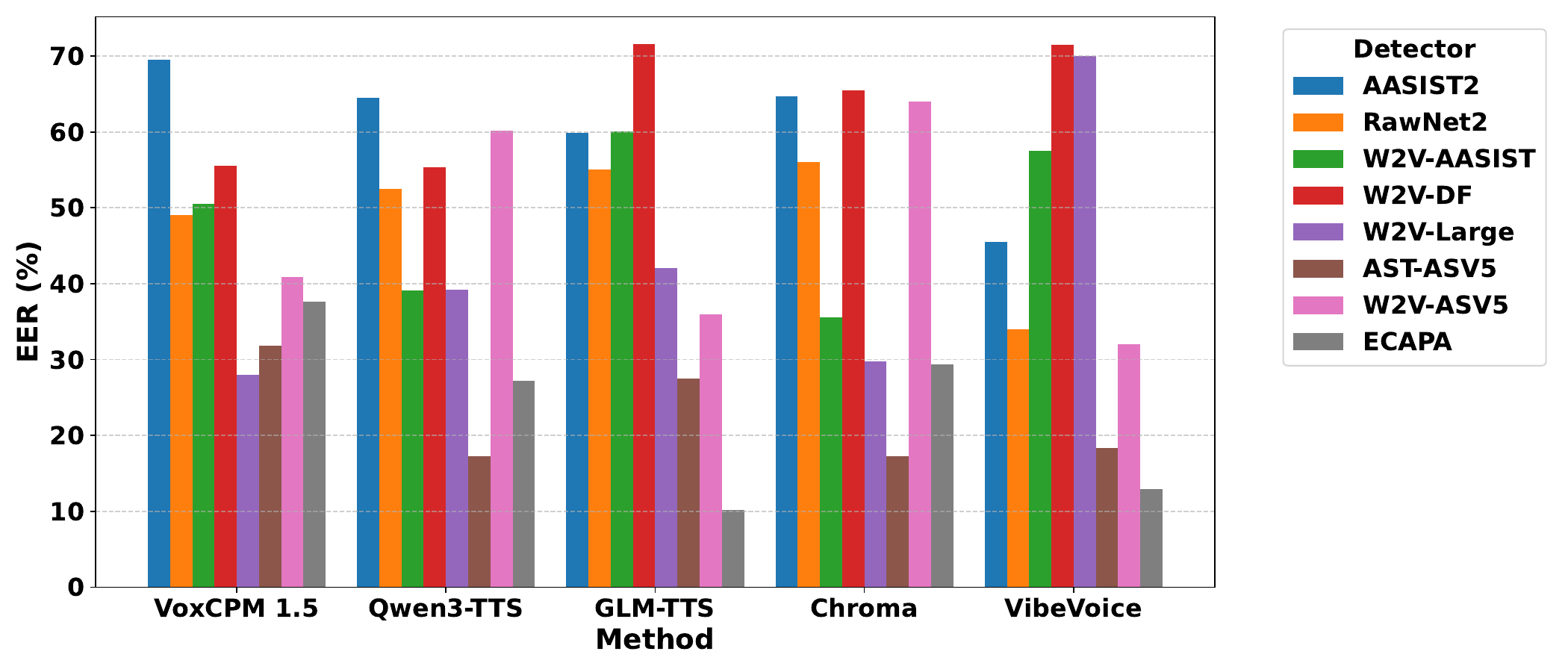}
    \includegraphics[width=0.49\textwidth]{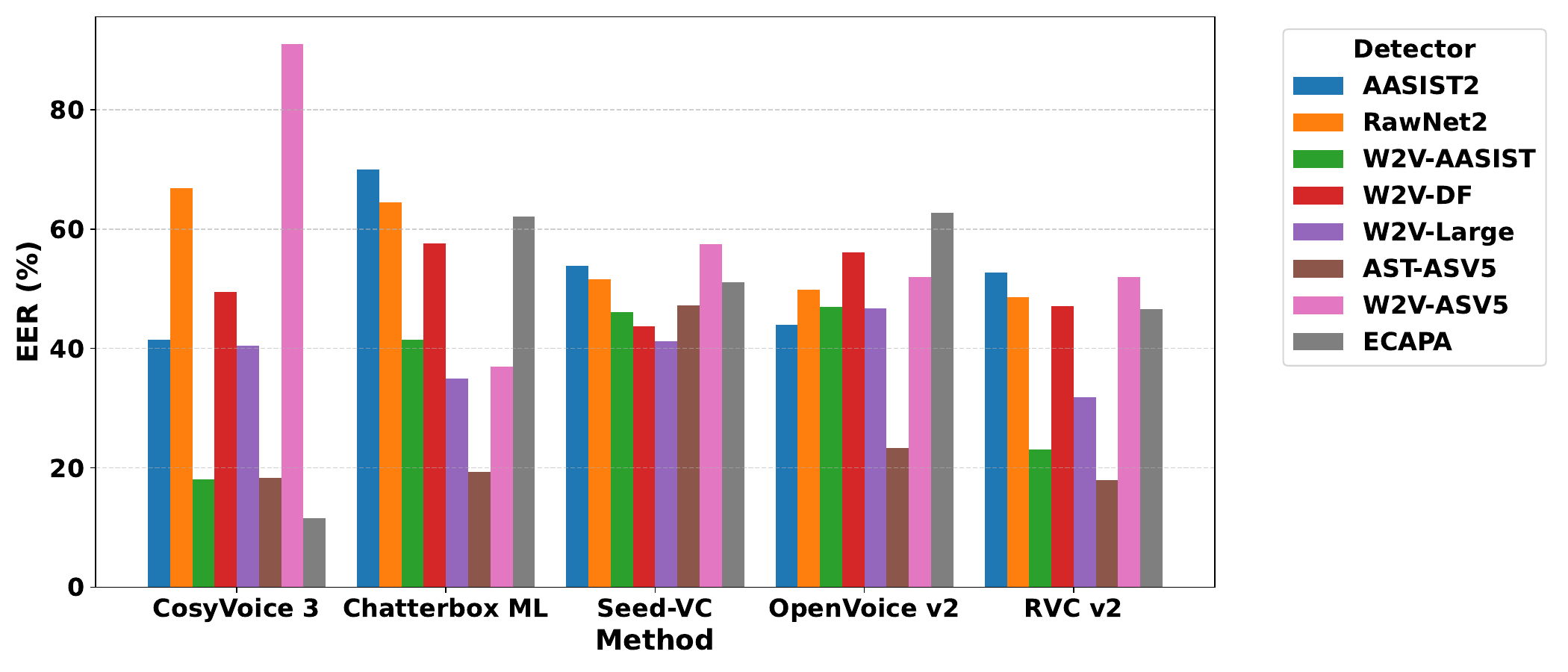}
    \caption{Per-method EER on Original Samples}
    \label{fig:per_method_eer}
    \vspace{-15pt}
\end{figure}

We evaluate different detectors on each speech synthesis method. As shown in Figure~\ref{fig:per_method_eer}, Seed-VC is the most challenging synthesis method overall: no evaluated detector achieves an EER below 41\%. We hypothesize that its diffusion-based conversion pipeline, augmented with strong speech representations (e.g., Whisper- and WavLM-derived features), yields highly natural outputs that preserve many bonafide time--frequency characteristics, thereby reducing detector-accessible artifacts. More broadly, no single detector is consistently reliable across all synthesis methods. For example, ECAPA-TDNN performs well on several TTS systems (e.g., GLM-TTS at 10.2\% and CosyVoice 3 at 11.5\%) but degrades substantially on VC (e.g., OpenVoice v2 at 62.7\%). The results suggest that speaker-embedding approaches may capture anomalies introduced by certain TTS pipelines yet struggle when VC more faithfully preserves speaker identity and natural speech structure. A promising future direction is to explore alternative modalities, such as acoustic sensing data~\cite{wang2023facer}, for deepfake audio detection.

\section{Conclusion}
We presented VoxENES 2026, a modern bilingual benchmark for speech spoofing detection that reflects LLM-era TTS and VC generation and realistic post-processing. Using VoxENES 2026, we evaluated eight pretrained detectors without fine-tuning and observed a substantial generalization gap: the best model achieves only 28.98\% EER overall, while many detectors perform near or below chance.
These results suggest that many current countermeasures rely on brittle, benchmark-specific artifacts and remain sensitive to modern generators and routine post-processing. We encourage future work on deployment-oriented spoofing detection that improves generalization under data drift and advances evaluation protocols that continuously track the fast-evolving TTS and VC frontier.

\newpage

\section{Use of Generative AI Disclosure}
Generative AI tools were used for language editing and manuscript polishing. All authors reviewed, verified, and take full responsibility for the content, experiments, and conclusions presented in this paper.

\bibliographystyle{IEEEtran}
\bibliography{mybib}

@inproceedings{jung2022aasist,
  title={{AASIST}: A new end-to-end anti-spoofing system using integrated spectro-temporal graph attention networks},
  author={Jung, Jee-weon and Heo, Hee-Soo and Tak, Hemlata and Shim, Hye-jin and Chung, Joon Son and Lee, Bong-Jin and Yu, Ha-Jin and Evans, Nicholas},
  booktitle={Proc. ICASSP},
  pages={6367--6371},
  year={2022}
}

@inproceedings{tak2021endtoend,
  title={End-to-end spectro-temporal graph attention networks for speaker verification anti-spoofing and speech deepfake detection},
  author={Tak, Hemlata and Jung, Jee-weon and Patino, Jose and Kamble, Madhu and Todisco, Massimiliano and Evans, Nicholas},
  booktitle={Proc. 2021 Edition of the Automatic Speaker Verification and Spoofing Countermeasures Challenge},
  pages={1--8},
  year={2021}
}

@inproceedings{desplanques2020ecapa,
  title={{ECAPA-TDNN}: Emphasized channel attention, propagation and aggregation in {TDNN} based speaker verification},
  author={Desplanques, Brecht and Thienpondt, Jenthe and Demuynck, Kris},
  booktitle={Proc. Interspeech},
  pages={3830--3834},
  year={2020}
}

@inproceedings{gong2021ast,
  title={{AST}: Audio Spectrogram Transformer},
  author={Gong, Yuan and Chung, Yu-An and Glass, James},
  booktitle={Proc. Interspeech},
  pages={571--575},
  year={2021},
  doi={10.21437/Interspeech.2021-698}
}

@inproceedings{todisco2019asvspoof,
  title={{ASVspoof} 2019: Future horizons in spoofed and fake audio detection},
  author={Todisco, Massimiliano and Wang, Xin and Vestman, Ville and Sahidullah, Md and Delgado, H{\'e}ctor and Nautsch, Andreas and Yamagishi, Junichi and Evans, Nicholas and Kinnunen, Tomi and Lee, Kong Aik},
  booktitle={Proc. Interspeech},
  pages={1008--1012},
  year={2019}
}

@article{yamagishi2021asvspoof,
  title={{ASVspoof} 2021: accelerating progress in spoofed and deepfake speech detection},
  author={Yamagishi, Junichi and Wang, Xin and Todisco, Massimiliano and Sahidullah, Md and Patino, Jose and Nautsch, Andreas and Liu, Xuechen and Lee, Kong Aik and Kinnunen, Tomi and Evans, Nicholas and Delgado, H{\'e}ctor},
  journal={IEEE/ACM Trans. Audio, Speech, and Language Processing},
  year={2024},
  publisher={IEEE}
}

@article{wang2026asvspoof5,
  title={{ASVspoof 5}: Design, collection and validation of resources for spoofing, deepfake, and adversarial attack detection using crowdsourced speech},
  author={Wang, Xin and Delgado, H{\'e}ctor and Tak, Hemlata and Jung, Jee-weon and Shim, Hye-jin and Todisco, Massimiliano and Kukanov, Ivan and Liu, Xuechen and Sahidullah, Md and Kinnunen, Tomi and Evans, Nicholas and Lee, Kong Aik and Yamagishi, Junichi},
  journal={Computer Speech \& Language},
  volume={95},
  year={2026},
  publisher={Elsevier}
}

@inproceedings{frank2021wavefake,
  title={{WaveFake}: A data set to facilitate audio deepfake detection},
  author={Frank, Joel and Sch{\"o}nherr, Lea},
  booktitle={Proc. NeurIPS Datasets and Benchmarks Track},
  year={2021}
}

@article{muller2022inthewild,
  title={Does audio deepfake detection generalize?},
  author={M{\"u}ller, Nicolas M and Czempin, Pavel and Dieckmann, Franziska and Frober, Adam and B{\"o}ttinger, Konstantin},
  journal={arXiv preprint arXiv:2203.16263},
  year={2022}
}

@inproceedings{muller2024mlaad,
  title={{MLAAD}: The Multi-Language Audio Anti-Spoofing Dataset},
  author={M{\"u}ller, Nicolas M and Kawa, Piotr and Choong, Wei Herng and Casanova, Edresson and G{\"o}lge, Eren and M{\"u}ller, Thorsten and Syga, Piotr and Sperl, Philip and B{\"o}ttinger, Konstantin},
  booktitle={Proc. IJCNN},
  year={2024}
}

@article{yan2024voicewukong,
  title={{VoiceWukong}: Benchmarking deepfake voice detection},
  author={Yan, Ziwei and Zhao, Yanjie and Wang, Haoyu},
  journal={arXiv preprint arXiv:2409.06348},
  year={2024}
}

@article{zhou2025voxcpm,
  title={{VoxCPM}: Tokenizer-free {TTS} for context-aware speech generation and true-to-life voice cloning},
  author={Zhou, Yixuan and Zeng, Guoyang and Liu, Xin and Li, Xiang and Yu, Renjie and Wang, Ziyang and Ye, Runchuan and Sun, Weiyue and Gui, Jiancheng and Li, Kehan and others},
  journal={arXiv preprint arXiv:2509.24650},
  year={2025}
}

@article{hu2026qwen3,
  title={{Qwen3-TTS} Technical Report},
  author={Hu, Hangrui and Zhu, Xinfa and He, Ting and Guo, Dake and Zhang, Bin and Wang, Xiong and Guo, Zhifang and Jiang, Ziyue and Hao, Hongkun and Guo, Zishan and others},
  journal={arXiv preprint arXiv:2601.15621},
  year={2026}
}

@article{cui2025glm,
  title={{GLM-TTS} technical report},
  author={Cui, Jiayan and Yang, Zhihan and Li, Naihan and Tian, Jiankun and Ma, Xingyu and Zhang, Yi and Chen, Guangyu and Yang, Runxuan and Cheng, Yuqing and Zhou, Yizhi and others},
  journal={arXiv preprint arXiv:2512.14291},
  year={2025}
}

@article{chen2026flashlabs,
  title={{FlashLabs Chroma 1.0}: A Real-Time End-to-End Spoken Dialogue Model with Personalized Voice Cloning},
  author={Chen, Tanyu and Chen, Tairan and Shen, Kai and Bao, Zhenghua and Zhang, Zhihui and Yuan, Man and Shi, Yi},
  journal={arXiv preprint arXiv:2601.11141},
  year={2026}
}

@inproceedings{pengvibevoice,
  title={{VibeVoice}: Expressive Podcast Generation with Next-Token Diffusion},
  author={Peng, Zhiliang and Yu, Jianwei and Wang, Wenhui and Chang, Yaoyao and Sun, Yutao and Dong, Li and Zhu, Yi and Xu, Weijiang and Bao, Hangbo and Wang, Zehua and others},
  booktitle={Proc. ICLR},
  year={2025}
}

@article{du2025cosyvoice3,
  title={{CosyVoice 3}: Towards in-the-wild speech generation via scaling-up and post-training},
  author={Du, Zhihao and Gao, Changfeng and Wang, Yuxuan and Yu, Fan and Zhao, Tianyu and Wang, Hao and Lv, Xiang and Wang, Hui and Shi, Xian and An, Keyu and others},
  journal={arXiv preprint arXiv:2505.17589},
  year={2025}
}

@misc{resemble2025chatterbox,
  title={Chatterbox: Open-Source Flow-Matching {TTS} with Emotion Exaggeration Control},
  author={{Resemble AI}},
  year={2025},
  howpublished={\url{https://huggingface.co/resemble-ai/chatterbox}}
}

@article{liu2024seedvc,
  title={Zero-shot Voice Conversion with Diffusion Transformers},
  author={Liu, Songting and others},
  journal={arXiv preprint arXiv:2411.09943},
  year={2024}
}

@misc{myshell2024openvoicev2,
  title={{OpenVoice V2}},
  author={{MyShell AI}},
  year={2024},
  howpublished={\url{https://huggingface.co/myshell-ai/OpenVoiceV2}}
}

@misc{rvcproject2024rvcwebui,
  title={Retrieval-based-Voice-Conversion-WebUI ({RVC})},
  author={{RVC-Project}},
  year={2024},
  howpublished={\url{https://github.com/RVC-Project/Retrieval-based-Voice-Conversion-WebUI}}
}

@inproceedings{panayotov2015librispeech,
  title={{LibriSpeech}: An {ASR} corpus based on public domain audio books},
  author={Panayotov, Vassil and Chen, Guoguo and Povey, Daniel and Khudanpur, Sanjeev},
  booktitle={Proc. ICASSP},
  pages={5206--5210},
  year={2015}
}

@inproceedings{wang2021voxpopuli,
  title={{VoxPopuli}: A large-scale multilingual speech corpus for representation learning, semi-supervised learning and interpretation},
  author={Wang, Changhan and Riviere, Morgane and Lee, Ann and Wu, Anne and Talber, Chaitanya and Bhosale, Jiatao and others},
  booktitle={Proc. ACL},
  pages={993--1003},
  year={2021}
}

@inproceedings{wang2023facer,
  title={FacER: Contrastive attention based expression recognition via smartphone earpiece speaker},
  author={Wang, Guangjing and Yan, Qiben and Patrarungrong, Shane and Wang, Juexing and Zeng, Huacheng},
  booktitle={IEEE INFOCOM 2023-IEEE conference on computer communications},
  pages={1--10},
  year={2023},
  organization={IEEE}
}

@inproceedings{wang2025clearmask,
  title={ClearMask: Noise-Free and Naturalness-Preserving Protection Against Voice Deepfake Attacks},
  author={Wang, Yuanda and Chen, Bocheng and Guo, Hanqing and Wang, Guangjing and Ding, Weikang and Yan, Qiben},
  booktitle={Proceedings of the 20th ACM Asia Conference on Computer and Communications Security},
  pages={696--709},
  year={2025}
}

@inproceedings{guo2024wavepurifier,
  title={WavePurifier: Purifying audio adversarial examples via hierarchical diffusion models},
  author={Guo, Hanqing and Wang, Guangjing and Chen, Bocheng and Wang, Yuanda and Zhang, Xiao and Chen, Xun and Yan, Qiben and Xiao, Li},
  booktitle={Proceedings of the 30th Annual International Conference on Mobile Computing and Networking},
  pages={1268--1282},
  year={2024}
}

@inproceedings{guo2023phantomsound,
  title={Phantomsound: Black-box, query-efficient audio adversarial attack via split-second phoneme injection},
  author={Guo, Hanqing and Wang, Guangjing and Wang, Yuanda and Chen, Bocheng and Yan, Qiben and Xiao, Li},
  booktitle={Proceedings of the 26th International Symposium on Research in Attacks, Intrusions and Defenses},
  pages={366--380},
  year={2023}
}

@inproceedings{wang2023vsmask,
  title={Vsmask: Defending against voice synthesis attack via real-time predictive perturbation},
  author={Wang, Yuanda and Guo, Hanqing and Wang, Guangjing and Chen, Bocheng and Yan, Qiben},
  booktitle={Proceedings of the 16th ACM Conference on Security and Privacy in Wireless and Mobile Networks},
  pages={239--250},
  year={2023}
}

\end{document}